\documentstyle[12pt,cites,epsf]{article}
\begin{document}
\renewcommand{\refname}{\protect\normalsize\protect\bf REFERENCES}
\numcite\bibnum{p}

\noindent
{\normalsize Published in {\em Physics of Manganites}, edited by T.A. Kaplan
 and   S. D. Mahanti, (Kluwer, New York, 1999), p. 103.}


\vspace {3.8cm}\noindent
{\bf OPTICAL CONDUCTIVITY OF DOPED MANGANITES:   }

\vspace{3mm}\noindent
{\bf COMPARISON OF FERROMAGNETIC KONDO LATTICE MODELS}

\vspace{3mm}\noindent
{\bf WITH AND WITHOUT ORBITAL DEGENERACY.}

\vspace{1.3cm}\hspace{2.0cm}
Frank Mack and Peter Horsch 

\vspace{4mm}\hspace{2.0cm}
Max-Planck-Institut f\" ur Festk\" orperforschung,

\hspace{2.0cm} Heisenbergstr. 1

\hspace{2.0cm} D-70569 Stuttgart, Germany.

\vspace{1.1cm}\noindent
{\bf INTRODUCTION}                             

\vspace{5mm}
The colossal magnetoresistance (CMR) of manganese oxides\cite{CMR}
and their transport properties
are usually studied in the framework of the double exchange (DE) Hamiltonian or the
more general ferromagnetic Kondo lattice model (KLM)\cite{Zener51} . 
The essence of these
models is the high-spin configuration of $e_g$-electron and $t_{2g}$-core
electron spins due to a strong ferromagnetic Kondo exchange interaction
$K\sim 1$eV. The kinetic energy in the partially filled $e_g$ band is lowered
when neighboring spins are aligned leading to a low-temperature ferromagnetic 
phase, while the high-$T$ paramagnetic phase is disordered with a high
resistivity. In these considerations the orbital degeneracy of the Mn $e_g$
orbitals is usually neglected.
Although this model provides an explanation of the
phases necessary for a qualitative understanding of CMR, it has been 
stressed that the double exchange mechanism is 
not sufficient  for a quantitative description\cite{Millis95}.

\begin{figure}
\epsfxsize=10cm
\centerline{\epsffile[60 20 370 281]{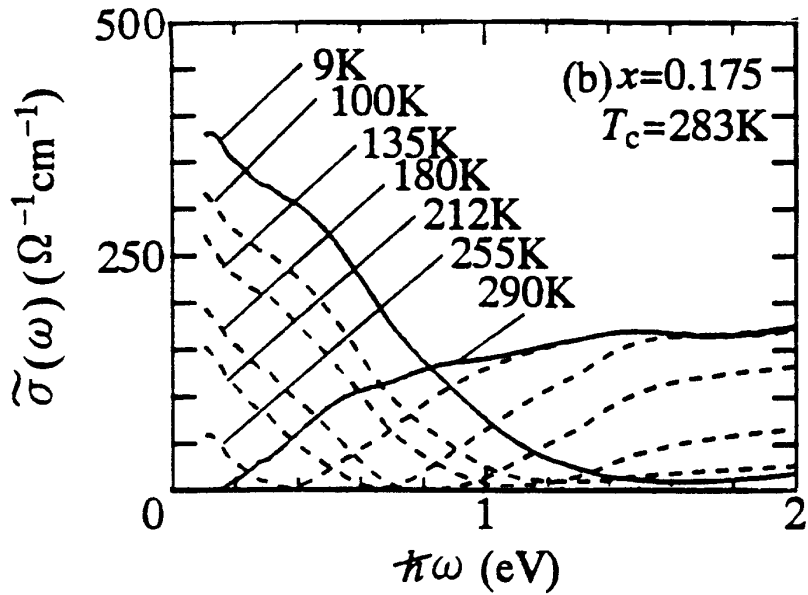}}
\noindent
\caption{\label{okimoto}
Temperature dependence of the optical conductivity of La$_{1-x}$Sr$_x$MnO$_3$
for $x=0.175$ (reproduced from Okimoto {\it et al.}). A temperature independent
background has been subtracted from the experimental data. 
}
\end{figure}

Doped manganites are characterized by strong correlations and the
complex interplay of spin-, charge-,
and orbital-degrees of freedom as well as the coupling to the lattice, e.g. via 
Jahn-Teller coupling \cite{Millis98,Ramirez97}.
This complexity is directly evident from the large number of phases in
a typical phase diagram.
To quantify the different mechanisms it is helpful to analyze experiments where 
for certain parameters one or the other degree of freedom is frozen out.
Important experiments in this respect are the very detailed investigations 
of the optical conductivity of La$_{1-x}$Sr$_x$MnO$_3$ 
by Okimoto {\it et al.}\cite{Okimoto95,Okimoto97}.
These experiments (see Fig. \ref{okimoto}) show (a) a pseudogap in $\sigma(\omega)$ 
for temperatures
above the Curie temperature $T_c$  (paramagnetic phase)
with  $\sigma(\omega)$ essentially linear in $\omega$, and
(b) the evolution of a broad incoherent distribution in the range
$0\leq \omega < 1.0 $ eV  below $T_c$, 
which still grows at temperatures below $T_c/10$,
where the magnetization is already close to saturation. 
Such a temperature dependence of $\sigma(\omega)$ over a wide energy region is
quite unusual as compared with other strongly correlated electron systems
near a metal-insulator phase boundary\cite{Okimoto95}. Interestingly
there is in addition a narrow Drude peak with width of about $0.02$ eV and
little weight superimposed to the broad incoherent spectrum. 
That is, the motion of charge carriers is dominated by incoherent processes
but there is also a coherent channel leading to a narrow Drude peak.
In view of the large energy scale of 1 eV it is plausible that the orbital 
degeneracy is the source of this incoherent motion, since the spin degrees of 
freedom are essentially frozen out. Our study of the degenerate Kondo
lattice model will support this point of view.
Further optical studies for various 3D manganites\cite{Kaplan96,Quijada98,Kim98}
as well as for the bilayer system La$_{1.2}$Sr$_{1.8}$Mn$_2$O$_7$
\cite{Ishikawa98} confirm the presence of the large 
incoherent absorption in the ferromagnetic state.

Although the importance of these experiments was recognized immediately,
the few theoretical studies\cite{Furukawa95,Ishihara97b,Shiba97} 
were confined to simplified models or approximations, 
thereby ignoring important aspects of the full many body problem.

The aim of this work is to show that in a model which accounts for the orbital 
degeneracy, yet assumes that the spins are fully polarized, the broad incoherent
spectral distribution of $\sigma(\omega)$, its increase with decreasing
temperature,
as well as the order of magnitude of  $\sigma(\omega)$ at small $\omega$
can be explained.
Our calculation also accounts for a small and narrow Drude peak as observed
by Okimoto {\it et al.} in the range $\omega<0.05$ eV. This is a clear indication
of coherent motion of charge carriers with small spectral weight, i.e.
in a model where due to the orbital degeneracy incoherent motion is dominant.

We start our discussion with a generic Hamiltonian for the manganite
systems, i.e. the ferromagnetic Kondo lattice model with degenerate
$e_g$-orbitals, and derive for the spin-polarized case an effective
model which contains only the orbital degrees of freedom. This orbital
model consists of a hopping term between the same and different
orbitals $\alpha$ and $\beta$ on neighbor sites
and an orbital interaction. Renaming  $\alpha=\sigma$ where $\sigma=
\uparrow$ or $\downarrow$ the model maps on a generalized anisotropic
$t$-$J$ model. The usual $t$-$J$ model known from the cuprates appears
as a special case. 
Our derivation includes the 3-site hopping processes, which appear as a natural
consequence of the strong coupling limit.
Although such terms do not influence the orbital order for integer filling,
they are important for the proper description of transport properties in the
doped systems\cite{Szczepanski90,Stephan92,Horsch93,Eskes94}.

Because of the complexity of the orbital model we present here a numerical
study of the orbital correlations and of the frequency dependent 
conductivity\cite{Horsch98}.
Although there exist studies of the interplay of orbital and spin
order at integer filling for LaMnO$_3$\cite{Ishihara97a,Feiner98},
the effect of doping has not been considered so far.
The finite temperature diagonalization\cite{Jaklic94,Jaklic98a} 
serves here as an unbiased tool to study the optical conductivity.
The change of orbital correlations as function of doping and temperature
is investigated by means of an {\em optimized orbital basis} (OOB)\cite{Horsch98}.

Furthermore we present results for the KLM {\it without} orbital
degeneracy which clearly show that this model
fails to explain the broad incoherent $\sigma(\omega)$
spectra observed in the ferromagnetic phase as $T\rightarrow 0$
\cite{Okimoto95,Okimoto97}. The KLM would instead lead to a sharp Drude
absorption because of the ferromagnetic alignement\cite{Jaklic98},
i.e. to spinless fermion behavior in the low-temperature limit.
This underlines the importance of the $e_g$ orbital degeneracy,
although there are alternative proposals invoking strong electron-phonon
coupling and lattice polaron formation\cite{Millis98,Millis95,Roeder96}
to explain the incoherency of the absorption.


\vspace{1.1cm}\noindent
{\bf MODELS}\newline

We start from the generic ferromagnetic Kondo lattice 
model for the manganites, where
the Mn $e_g$-electrons are coupled to core spins
$\vec{S}_{\bf i}$ (formed by the $t_{2g}$ orbitals) 
and a local repulsion $U$ between the electrons in the two
$e_g$ orbitals:
\begin{equation}
H = H_{band} + H_{int} + H_{Kondo}.
\label{eq:H}
\end{equation}
$H_{band}$ describes the hopping of $e_g$-electrons  between
sites $i$ and $j$, $H_{int}$ the interaction between $e_g$ electrons and
$H_{Kondo}$ the coupling of $e_g$-spins and $t_{2g}$-core spins.  
The $e_g$ electrons to have two-fold orbital degeneracy labelled
by a Roman index (a,b) and two fold spin degeneracy
labelled by a Greek index $( \sigma , \sigma' )$.  Explicitly,
\begin{equation}
H_{band}= \sum_{{\bf i} a\sigma} E_{\bf i}^a
d_{{\bf i} a\sigma}^{\dagger} d_{{\bf i} a\sigma} 
+ \sum_{\langle {\bf i j} \rangle ab\sigma} (t_{{\bf i j}}^{ab}
d_{{\bf i} a \sigma}^{\dagger} d_{{\bf j} b \sigma}+ H.c.).
\label{eq:Hband}
\end{equation}
The hopping matrix elements $t_{{\bf ij}}^{ab}$ form a real
symmetric matrix whose form depends on the choice
of basis in orbital space and the direction of the ${\bf i-j}$ bond.
For the present study we shall use $|x\rangle\sim x^2-y^2$ and 
$|z\rangle \sim (3z^2-r^2)/\sqrt{3}$ 
as basis for the $e_g$ orbitals. 
From the Slater-Koster rules follows\cite{Slater54,Kugel73,Harrison80}:
\begin{equation}
t_{{\bf ij}\parallel x/y}^{ab}=-\frac{t}{4}\left( \begin{array}{cc}
1 & \mp\sqrt{3}\\
\mp\sqrt{3} & 3\\
\end{array} \right),\;\;
t_{{\bf ij}\parallel z}^{ab}=-t\left( \begin{array}{cc}
1 & 0\\
0 & 0\\
\end{array} \right),
\end{equation}
which allows for inter-orbital hopping in the $xy$-plane, where the upper
(lower) sign distinguishes hopping along $x$ and $y$ direction.
The hopping matrix elements are defined in terms of the 
double exchange $t$ of $d_{3z^2-r^2}$ orbitals along the $c$-axis.
Here $t={V^2_{dp\sigma}}/{\Delta\epsilon_{dp}}$
is determined by the Mn-$d$ O-$p$ hybridisation $V_{dp\sigma}$ and the
corresponding level splitting ${\Delta\epsilon_{dp}}$. 
All matrix elements are negative except $t^{xz}$ along x-direction,
which is positive. 
The level splitting $\Delta E_{\bf i}=E_{\bf i}^z - E_{\bf i}^x$ can be controlled
by uniaxial pressure or in the case of layered compounds even with
hydrostatic pressure\cite{Ishihara98c}.
We shall focus here on the orbital degenerate case,  
i.e.  $\Delta E_{\bf i}=0$, however we will keep
the orbital energy term for the derivation of the orbital model.

The large local electron-electron repulsion $U$ is 
responsible for the insulating behavior in the case of integer band-filling
\begin{equation}
H_{int} = \sum_{{\bf i} a} U_{a} n_{{\bf i} a\uparrow}n_{{\bf i} a\downarrow}+
    U_{ab} \sum_{\bf i} n_{{\bf i} a}n_{{\bf i} b} 
+ J_{ab}\sum_{{\bf i}\sigma \sigma'} {d}^{\dagger}_{{\bf i} a \sigma}
     {d}^{\dagger}_{{\bf i} b \sigma'} {d}_{{\bf i} a \sigma'}
      {d}_{{\bf i} b \sigma}.
\label{eq:H_{dint}}
\end{equation}
Here $U_a$, $U_{ab}$ and $J_{ab}$ are the intra- and inter-orbital Hubbard and
exchange interactions, respectively.
The relevant valences of Mn are Mn$^{3+}$ (S=2) and  Mn$^{4+}$
(S=3/2) as was pointed out already by Zener\cite{Zener51}.
The Hamiltonian $H_{band}+H_{int}$ was considered
     by Zaanen and Ole\'s who showed that in general rather complex
     effective Hamiltonians result for transition metal ions with 
     partially filled $d$ shell near orbital degeneracy \cite{Zaanen93}.
     Here a simplified approach is preferred, with the $t_{2g}$ electrons
     of Mn ions forming core spins of $S=3/2$, and thus we restrict the
     electron interactions in $H_{int}$ to the $e_g$ bands.

Even in the case with local orbital degeneracy there is one lower
Hubbard band which is partially filled in the case of hole doping.
With one $e_g$-electron per site these systems are Mott-insulators. 
Although the Jahn-Teller splitting
can lead to a gap in the absence of $U$, it is not the primary
reason for the insulating behavior for integer filling\cite{Varma96}. 

The interaction of itinerant electrons
with the (S=3/2) core spins is given by
\begin{equation}
H_{Kondo} = -K\sum_{{\bf i} a\sigma\sigma'} \vec{S}_{\bf i} \cdot
d_{{\bf i} a\sigma}^{\dagger}\vec{\sigma}_{\sigma\sigma'}
d_{{\bf i} a\sigma'}
\label{eq:H_{dex}}
\end{equation}
leading to parallel alignment of the d-electron spin with the core-spin
$\vec{S}_{\bf i}$. Since the d-electron kinetic energy is favored by a
parallel orientation of neighboring spins the ground state becomes
ferromagnetic\cite{Zener51}.

In the low-temperature ferromagnetic phase we may introduce an
effective Hamiltonian which contains only the orbital degrees of
freedom assuming a fully spin-polarized ferromagnet.
The spin degrees of freedom can also be eliminated by high magnetic
fields. This opens the possibility to investigate the orbital order
independent of the spin degrees of freedom, and to shed light
on the nontrivial
question how the orbital order is changed upon doping with
$e_g$-electrons or holes.

The resulting model has similarities to the $t$-$J$ model, where the
spin-indices $\sigma$ and $\sigma'$ are now orbital indices. To stress
the difference we use for the orbital indices the letters $a,b$ and
$\alpha, \beta$. Unitary transformation $H_{orb}=e^{-S}He^S$
\cite{MacDonald88} and 
the restriction to states without double occupancy leads to the
{\it orbital} $t$-$J$ model which has the following structure:
\begin{equation}
H_{orb}= \sum_{{\bf i} a} E^{\bf i}_a
\tilde{d}_{{\bf i} a}^{\dagger} \tilde{d}_{{\bf i} a}
 + \sum_{\langle {\bf ij} \rangle ab} (t_{\bf ij}^{ab}
\tilde{d}_{{\bf i} a}^{\dagger} \tilde{d}_{{\bf j} b}+ H.c.)+H'_{orb}.
\label{eq:Hfm}
\end{equation}
with the constraint that each site can be occupied by at most one
electron, i.e. 
${\tilde d}^{\dagger}_{{\bf i}a}=d^{\dagger}_{{\bf i}a}(1-n_{{\bf i}{\bar a}})$.
Here and in the following the index ${\bar a}$ denotes the orthogonal $e_g$ orbital
with respect to orbital $a$.
The orbital interaction $H'_{orb}$ follows as a consequence of the 
elimation of doubly occupied sites with energy $U\sim U_{ab}-J_{ab}$.\cite{Horsch98} 
\begin{eqnarray}
H'_{orb}  =  -\frac{1}{2}\sum_{\bf j u u'}\sum_{a b \alpha \beta}
t^{\alpha \beta}_{\bf j+u\; j}t^{b a}_{\bf j \; j+u'}
\biggl( \frac{1}{ U+E_{\bf j}^{\beta}-E_{\bf j+u}^{\alpha} }  
+\frac{1}{ U+E_{\bf j}^b-E_{\bf j+u'}^a } & \biggr)& \nonumber\\ 
 \Bigl[ \delta_{\beta,b}\; \tilde{d}^{\dagger}_{{\bf j+u} \alpha}
\tilde{d}^{\dagger}_{{\bf j} \bar{\beta}} \tilde{d}_{{\bf j} \bar{b}}
\tilde{d}_{{\bf j+u'} a} 
 -\delta_{\beta,\bar{b}}\; \tilde{d}^{\dagger}_{{\bf j+u} \alpha}
\tilde{d}^{\dagger}_{{\bf j} b} \tilde{d}_{{\bf j} \beta}
\tilde{d}_{{\bf j+u'} a}  &\Bigr]& .
\label{eq:H_orb0}
\end{eqnarray}
Here ${\bf u, u'}=(\pm a,0)$ or $(0,\pm a)$ are lattice unit vectors.
The orbital interaction  $H'_{orb}=H'^{(2)}_{orb}+H'^{(3)}_{orb}$
consists of two types of
contributions: (i) 2-site terms
(${\bf u}={\bf u'}$), i.e. similar to the Heisenberg interaction in the
standard $t$-$J$ model, yet more complex because of the nonvanishing 
off-diagonal $t^{ab}$, and (ii) 3-site hopping terms
(${\bf u} \neq {\bf u'}$) between second nearest neighbors.
In the half-filled case, i.e. one $e_g$-electron per site, only the
2-site interaction is operative and may induce some orbital order.
In the presence of hole doping, i.e. for less than one $e_g$-electron
per site, both the kinetic energy and the 3-site contributions in
$H'_{orb}$ lead to propagation of the holes and to a frustration of
the orbital order.

The complexity of the model can be seen if we express the orbital
interaction $H'_{orb}$ in terms of pseudospin operators
$T_{\bf i}^z=\frac{1}{2}(n_{{\bf i}a}-n_{{\bf i}b})$,
$T_{\bf i}^+=\tilde{d}^{\dagger}_{{\bf i}a}\tilde{d}_{{\bf i}b}$
and
$T_{\bf i}^-=\tilde{d}^{\dagger}_{{\bf i}b}\tilde{d}_{{\bf i}a}$.
where $a=\uparrow$ and 
$b=\downarrow$ denote an orthogonal orbital basis in the
$e_g$-space. To be specific we shall assume here $a(b)=z(x)$.
Keeping only the two-site contributions, i.e.  ${\bf u}={\bf u'}$,
one obtains an anisotropic Heisenberg Hamiltonian for the orbital
interactions:
\begin{eqnarray}
H'^{(2)}_{orb}=-\frac{2}{U}\sum_{\langle {\bf i j} \rangle}&\Bigl[ &
\bigl({t_{\bf i j}^{a a}}^2+{t_{\bf i j}^{b b}}^2\bigr)
\bigl(\frac{1}{4}n_{\bf i}n_{\bf j}-T_{\bf i}^zT_{\bf j}^z\bigr) 
- t_{\bf i j}^{a a}t_{\bf i j}^{b b}
\bigl(T_{\bf i}^+T_{\bf j}^-+T_{\bf i}^-T_{\bf j}^+\bigr)\nonumber\\
&+&\bigl({t_{\bf i j}^{a b}}^2+{t_{\bf i j}^{b a}}^2\bigr)
\bigl(\frac{1}{4}n_{\bf i}n_{\bf j}+T_{\bf i}^zT_{\bf j}^z\bigr) 
- t_{\bf i j}^{a b}t_{\bf i j}^{b a}
\bigl(T_{\bf i}^+T_{\bf j}^++T_{\bf i}^-T_{\bf j}^-\bigr)\\
&-&\bigl(t_{\bf i j}^{a a}t_{\bf i j}^{a b}
 -t_{\bf i j}^{bb}t_{\bf i j}^{ba}\bigr)
\bigl(T_{\bf i}^zT_{\bf j}^++T_{\bf i}^zT_{\bf j}^-\bigr)
-\bigl(t_{\bf i j}^{a a}t_{\bf i j}^{ba}
 -t_{\bf i j}^{bb}t_{\bf i j}^{ab}\bigr)
\bigl(T_{\bf i}^+T_{\bf j}^z+T_{\bf i}^-T_{\bf j}^z\bigr)
\Bigr]. \nonumber 
\label{eq:H_orb}
\end{eqnarray}
The orbital interaction $H'^{(2)}_{orb}$ is equivalent to that studied
by Ishihara {\it et al.}\cite{Ishihara97a,Ishihara98c}. 
Since we are interested
here in the effect of holes on the orbital order we will generally
base our study on $H'_{orb}$ (\ref{eq:H_orb0}) which includes the 3-site hopping
processes as well.

We note that in the special case $t_{\bf i j}^{a b}=
\delta_{a b} t$ and $E_{\bf i}^a =0$ equations (\ref{eq:Hfm}) and (\ref{eq:H_orb0}) 
are identical to
the usual $t$-$J$ model ( with $J=4 t^2/U$ and including 3-site processes).
For the orbital model we shall adopt the same convention for the orbital
exchange coupling, i.e.  $J=4 t^2/U$.

The orbital degrees of freedom in combination with strong correlations
(i.e. no doubly occupied sites are allowed) is expected to lead to incoherent 
motion of holes, although quasiparticle formation
with reduced spectral weight is a plausible expectation
on the basis of what is known about the usual $t$-$J$ model.

In this paper we focus on the orbital degenerate
case with $E_i^a=E_i^b=0$. A detailed study of
the influence of a finite level splitting will be presented elsewhere.

\vspace{1.1cm}\noindent
{\bf OPTICAL CONDUCTIVITY}\newline

We study the charge transport by calculating the optical conductivity
\begin{equation}
\sigma_0(\omega)=2\pi e^2 D_c\delta(\omega)+\sigma(\omega).
\label{eq:sigma0}
\end{equation}
The frequency dependent conductivity consists of two parts, the regular
finite frequency absorption $\sigma(\omega)$ and the 
$\delta$-function contribution which is proportional to the charge
stiffness $D_c$\cite{Kohn64,Shastry90}  . 
The latter vanishes in insulators. This contribution is
broadened into a usual Drude peak in the presence of other scattering
processes like impurities which are not contained in the present
model. The finite frequency absorption (or regular part) $\sigma(\omega)$
is determined by the current-current correlation function using the Kubo
formula
\begin{equation}
\sigma(\omega)=\frac{1-e^{-\omega/T}}{N\omega} Re \int_0^{\infty}dt
e^{i\omega t}\langle j_x(t)j_x\rangle.
\label{eq:sigma}
\end{equation}
For the derivation of the current operator we introduce twisted
boundary conditions via Peierls construction
\begin{equation}
t_{\bf j+u \; j}^{a b}({\vec A})=t^{a b}_{\bf j+u \; j} 
\exp\biggl(-i \frac{e}{\hbar} 
\int_{\bf j}^{\bf j+u} {\vec A}(x){\vec {dx}}\biggr).
\label{eq:t_peierls}
\end{equation}
Assuming ${\vec A}=(A_x,A_y,0)$ we obtain from the kinetic energy
operator the 
x-component of the current operator $j_x=\partial H(A_x)/\partial A_x$:
\begin{equation}
j^{(1)}_x=-ie \sum_{{\bf j+u} a b} t^{a b}_{\bf j+u \; j}\; u_x \;
{\tilde d}^{\dagger}_{{\bf j+u} a} {\tilde d}_{{\bf j} b}.
\label{eq:j(1)}
\end{equation}
An additional contribution follows from the 3-site term in Eq.(6)
\begin{eqnarray}
j^{(3)}_x&=& \frac{ie}{2U} \sum_{\bf j u u'}\sum_{a b \alpha \beta}
t^{\alpha \beta}_{\bf j+u\; j}t^{b a}_{\bf j \; j+u'}(u_x-u_x')
O^{\alpha \beta}_{ba}(\bf j, u, u') ,\nonumber\\
\\
O^{\alpha \beta}_{ba}&=&
\delta_{\beta,b}\; \tilde{d}^{\dagger}_{{\bf j+u} \alpha}
\tilde{d}^{\dagger}_{{\bf j} \bar{\beta}} \tilde{d}_{{\bf j} \bar{b}}
\tilde{d}_{{\bf j+u'} a}  
-\delta_{\beta,\bar{b}}\; \tilde{d}^{\dagger}_{{\bf j+u} \alpha}
\tilde{d}^{\dagger}_{{\bf j} b} \tilde{d}_{{\bf j} \beta}
\tilde{d}_{{\bf j+u'} a}, \nonumber
\end{eqnarray}
where the expression $O^{\alpha \beta}_{ba}(\bf j, u, u')$ is an 
abbreviation for the operator within the last brackets of Eq.(6).

We stress here that the orbital order in the undoped case
is determined exclusively by the (two-site) orbital interaction (7),
while the 3-site processes in (6) only contribute in the doped
case. Nevertheless the motion of holes and transport in general
is influenced in a significant way by the 3-site current 
operator $j^{(3)}_x$. In fact from studies of the $t$-$J$ model it
is known that $\sigma(\omega)$ is not only quantitatively but
even qualitatively changed by these terms\cite{Stephan92,Horsch93}.
Although the 3-site terms are usually ignored in studies of the
$t$-$J$ model they are an important part of the strong coupling model
and should not be droped in studies of the conductivity. 
One of the aims in our study of the orbital model is to analyse the
effect of these 3-site terms in the model.

The charge stiffness $D_c$ can be determined in two ways: (a) using Kohn's 
relation\cite{Kohn64}, 
or (b) via the optical sum rule which relates the integrated
spectral weight of the real part of the conductivity to the average kinetic
energy:
\begin{equation}
\int^{\infty}_{-\infty} \sigma_0(\omega)d\omega=
-\frac{\pi e^2}{N}\langle H^{kin}_{xx}\rangle.
\label{eq:sumr}
\end{equation}
Here $H^{kin}$ contains two contributions: (a) the usual kinetic energy
$\sim t$ (5) and (b) the 3-site hopping processes $\sim t^2/U$ in Eq. (6).
The case with off-diagonal hopping considered here is a generalization
of the sum rules for the $t$-$J$ model\cite{Baeriswyl86} and for the $t$-$J$
model including 3-site terms\cite{Stephan92,Eskes94}.
Together with (8) this implies
\begin{equation}
D_c=-\frac{1}{2N}\langle H^{kin}_{xx}\rangle
    -\frac{1}{\pi e^2} \int^{\infty}_{0^+}\sigma(\omega)d\omega.
\label{eq:Dc}
\end{equation}
In the following we shall use $D_c=S-S_{\omega}$ as abbreviation for this
equation, where $S$ denotes the sum rule expression 
and $S_{\omega}$ the finite frequency absorption.

\vspace{1.1cm}\noindent
{\bf FINITE TEMPERATURE LANCZOS METHOD (FTLM)}\newline

For the calculation of $\sigma(\omega)$ we use a generalization of
the exact diagonalization technique for finite temperature developed
by Jakli\v c and Prelov\v sek\cite{Jaklic94}.  In this approach 
the trace of the thermodynamic expectation value is performed by a 
Monte-Carlo sampling.
The current-current correlation function in (9):
\begin{equation}
C(\omega)= Re \int_0^{\infty}dt e^{i\omega t}\langle j_x(t)j_x\rangle.
\label{eq:C1}\end{equation}
can be rewritten by introducing a complete set of basis functions 
$\mid r \rangle$ for the
trace, and eigenfunctions $\mid {\Psi}_{j}^r \rangle$ and  
$\mid \tilde{\Psi}_{j}^r \rangle$ of $H$ with eigenvalues ${E}^r_i$  
and $\tilde{E}^r_j$ (here $r$ is an irrelevant label, which will get
its meaning below):
\begin{equation}
C(\omega)\approx  \frac{\pi}{Z}\sum_{r=1}^{R}\sum_{i,j=1}^{M} e^{-\beta
  E_{i}^r}
 \langle  r \mid \Psi_i ^r \rangle 
 \langle \Psi_i ^r \mid j_x\mid \tilde{\Psi}_{j}^r \rangle 
\langle \tilde{\Psi}_{j}^r\mid j_x\mid r \rangle \delta(\omega+E_i ^r -\tilde{E}_j^r)
\label{eq:C2}\end{equation}
with 
\begin{equation}
Z \approx \sum_{r=1}^{R}\sum_{i=1}^{M} e^{-\beta E_i^r} \mid  \langle r \mid \Psi
_i^r \rangle \mid ^2 .
\label{eq:Z}\end{equation} 
These expressions are exact if one uses complete sets.
The FTLM is based on two approximations when evaluating Eqs.(\ref{eq:C2}) 
for $C(\omega)$ and (\ref{eq:Z}) for the partition function $Z$:
(1) The trace is performed over a restricted number $R$ of random 
states $\mid r \rangle$, and (2) the required set of eigenfunctions of
$H$ is generated by the Lanczos algorithm starting from the initial
states $  \mid {\Phi}_{0}^r \rangle =\mid r \rangle$ and 
$\mid \tilde{\Phi}_{0}^r \rangle
=j_x \mid r \rangle/\sqrt{ \langle  r \mid j_x^2 \mid r \rangle}$,
respectively.
The latter procedure is truncated after $M$ steps and yields the relevant
intermediate states for the evaluation of $C(\omega)$.

Detailed tests \cite{Jaklic94} have shown that the results get very accurate
already if $R,M \ll N_{0} $, where $N_{0}$ is the dimension of the
Hilbert space. 
The  computational effort is significantly reduced
because only matrices of dimension  $M\times M$ have to be
diagonalized, i.e. much smaller than the dimension 
$ N_{0} \times N_{0}$ of the full Hamiltonian $H$.
In practice typical dimensions are $R=100$ and $M=100$.

A complication in the case of the orbital $t$-$J$ model is that
$T^z_{tot}$ does not commute with the Hamiltonian $H_{orb}$ (\ref{eq:Hfm}). 
Therefore the calculations for the orbital model are restricted 
to smaller clusters than in the $t$-$J$ case, where different 
$S^z_{tot}$ subspaces can be treated separately.

\vspace{1.1cm}\noindent
{\bf RESULTS}\newline                  
\noindent
{\bf Kondo-Lattice Model without Orbital Degeneracy}\newline

We begin our discussion of the frequency dependent conductivity with
results for the {\it single orbital Kondo-lattice model} excluding double occupancy
\begin{equation}
H= -\sum_{{\bf ij}\sigma} t_{\bf ij}
{\tilde d}_{{\bf i}\sigma}^{\dagger} {\tilde d}_{{\bf j}\sigma}
-J_H\sum_{{\bf i}\sigma\sigma'} \vec{S}_{\bf i}\cdot
{\tilde d}_{{\bf i}\sigma}^{\dagger}\vec{\sigma}_{\sigma\sigma'}
{\tilde d}_{{\bf i}\sigma'}.
\label{eq:H_Kondo}
\end{equation}
Nearest neighbor hopping $ t_{\bf ij}=t$ is assumed
and the actual calculations were
performed for $S=1$ core spins for different doping concentrations $x$.

\begin{figure}
\epsfxsize=6cm
\centerline{\epsffile[60 20 530 680]{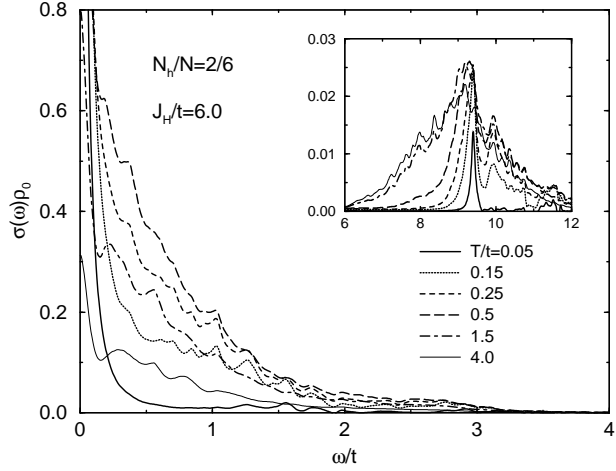}}
\caption{Frequency dependent conductivity $\sigma(\omega)$ of the {\it single
  orbital KLM} for a six
  site chain with two electrons and $J_H=6 t$ for different temperatures.
  Inset shows the spectrum of excitations to the exchange-splitted band.
  Note that for $T\rightarrow 0$ all spectral weight is in the Drude peak
  (From Jakli\v c, Horsch and Mack\cite{Jaklic98}).
\label{sigma_T}}
\end{figure}

\begin{figure} 
\epsfxsize=6cm
\centerline{\epsffile[60 20 530 680]{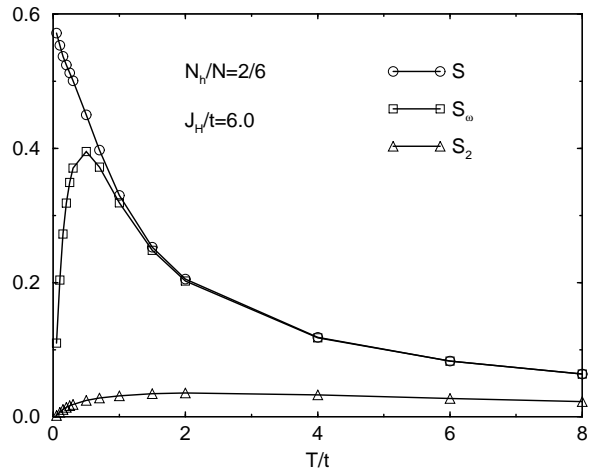}}
\caption{Temperature dependence of the kinetic energy and of the 
  optical spectral weights of the ferromagnetic {\em Kondo-lattice model 
  in the absence of orbital degeneracy} for a 6 site chain with two holes.
  Here 
  $S=\langle H^{kin}_{xx}\rangle/2Nt$ and 
  $S_{\omega}=\int \sigma(\omega)d\omega/\pi e^2t$. The spectral weight of 
  the interband optical transitions into low-spin states is measured by 
  $S_2=\int \sigma_2(\omega)d\omega/\pi e^2t$.
\label{sumrule}}
\end{figure}
In Fig.\ref{sigma_T} the temperature dependence of the optical conductivity
is shown for a chain at doping $x=2/6$. The conductivity is given in 
dimensionless form $\sigma\rho_0$, where  $\rho_0=\hbar a /e^2$ sets the dimension
of the resistivity in a 3D s.c. crystal structure with lattice constant $a$.
At strong coupling $J_H\gg t$ the states of the model consist of a low-energy
band corresponding to the $S+1/2$ configurations, and of the exchange-splitted bands 
separated by an energy $\frac{3}{2} J_H$. 
The intraband transitions are shown in the large window.
They give rise to an incoherent low-energy absorption extending up to the energy of 
the free carrier band-width $\sim 4t$.  The spectrum of the high to low spin
transitions is centered around $\omega\sim \frac{3}{2} J_H$ and is shown in the inset.

At low temperatures, however, when ferromagnetic correlations extend over the whole
system, the low-energy spectrum is dominated by a 
narrow coherent peak at $\omega=0$, and the interband transitions vanish.
At $T=0$ all weight is in the Drude peak (see Fig.\ref{sumrule}).
As the temperature is increased the spectral weight is transfered from the coherent 
part to the broad incoherent spectrum and to the exchange-split excitations.
This clearly shows that the model without orbital
degeneracy does not explain the anomalous absorption in the 
ferromagnetic state of manganites (Fig.\ref{okimoto}).\newline
\newline\noindent
{\bf Optical Conductivity for the Orbital $t$-$J$ Model}\newline

\begin{figure}
\epsfxsize=6.6cm
\centerline{\epsffile[60 20 530 680]{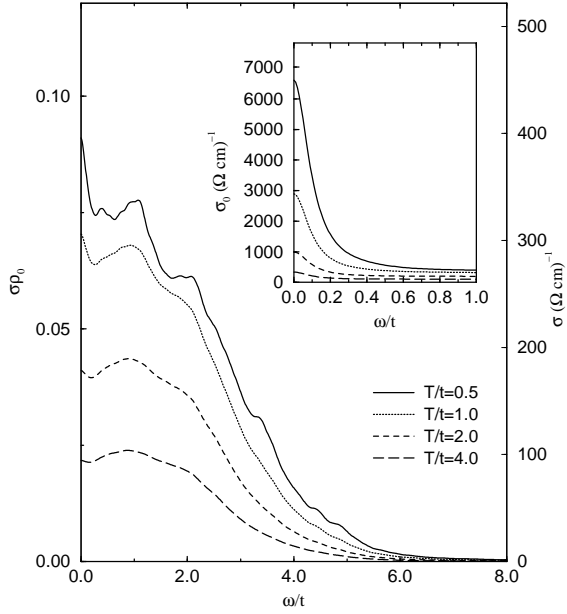}}
\noindent
\caption{ \label{fig1}
Optical conductivity $ \sigma(\omega) $ calculated  for the {\em orbital $t$-$J$ model}
for a N=10 site plane with $ N_e $=8  electrons and J=0.25t (U=16t)
 at different temperatures. The calculation includes the 3-site terms.
The inset shows the optical conductivity $ \sigma_0
(\omega) $ including the Drude-part. 
The spectra were broadened with $ \Gamma $=0.1t.  }
\end{figure}

\begin{figure}
\epsfxsize=6.6cm
\centerline{\epsffile[60 20 530 680]{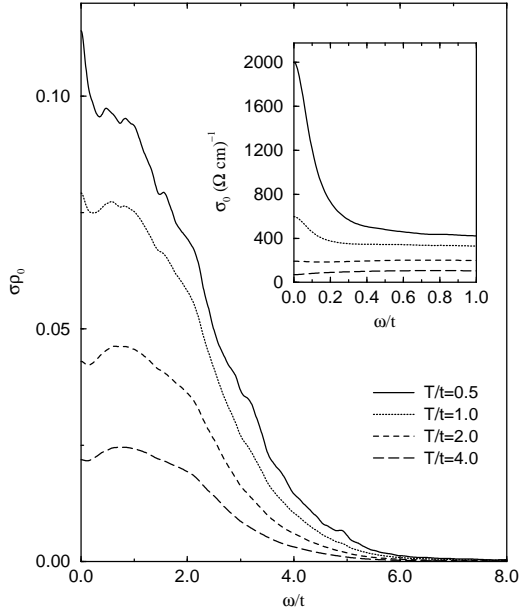}}
\noindent
\caption{ \label{fig2}
Optical conductivity as in Fig.\ref{fig1} but without 3-site contributions. Due
to the stronger incoherence there is a smaller Drude peak (inset)
and a smaller DC-conductivity.  (From Horsch, Jakli\v c and Mack\cite{Horsch98}).
}
\end{figure}

Typical results for the frequency and
temperature dependence of $\sigma(\omega)$
for a two-dimensional 10-site cluster with 2 holes, i.e. corresponding to a
doping concentration $x=0.2$, are shown in Fig. \ref{fig1} with and 
Fig. \ref{fig2} without 
3-site hopping terms, respectively.
The $\sigma(\omega)$ spectra are bell-shaped and increase with decreasing temperature.
The width of the distribution  measured at half-maximum is $\omega_{1/2}\sim 2.5 t$ 
and is essentially  independent
of temperature. Calculations for a 3D cluster yield  a rather
similar $\sigma(\omega)$ distribution with
a slightly larger value for the width  $\omega_{1/2}\sim 3.5 t$.

If we compare this latter (3D) value with $\omega^{exp}_{1/2}\sim 0.7$ eV 
found from the data of Okimoto {\it et al.} (for $x=0.175$ and $T=9$K),
we obtain an experimental estimate for 
the parameter $t$: $t^{opt}\sim 0.2$ eV.
For comparison a rough theoretical estimate based on Harrison's solid
state table yields $t=V^2_{dp\sigma}/\Delta\epsilon \sim 0.4$ eV. 
Hence the orbital $t$-$J$ model explains in a natural way the $\sim 1$ eV energy scale
of the incoherent absorption in the experiments.
Although $\sigma(\omega)$ shows the anomalous increase with decreasing temperature, 
we do not claim here that the orbital model accounts for the full temperature 
dependence in the ferromagnetic phase ($T<T_c$). The description of the
complete $T$-dependence certainly requires the analysis of the degenerate-orbital KLM,
i.e. the additional inclusion of the spin degrees of freedom. Yet we believe that the 
results can be compared with the data by Okimoto {\it et al.}
in the low-$T$ limit, i.e. in the regime where the magnetization is saturated. 

To compare with experimental data we present the conductivity in dimensionless
form and with the proper dimensions of a 3D conductivity, assuming that the 3D system 
is formed by noninteracting layers.  For a cubic lattice $\rho_0=\hbar a /e^2$.
If we consider La$_{1-x}$Sr$_x$MnO$_3$ with lattice constant $a=5.5\;\AA$ and 
$\hbar/e^2=4.11$K$\Omega$ we obtain $\rho_0=0.23\cdot 10^{-3}\; \Omega$cm.
Hence the low frequency limit of the regular part of the conductivity
$\sigma(\omega\rightarrow 0)=\sigma_0^{inc}\sim 0.3-0.4\cdot 10^3$ ($\Omega$cm)$^{-1}$.
This is consistent with the order of magnitude for the low frequency limit 
of the incoherent 
part of the $\sigma(\omega)$ data of Okimoto {\it et al.} for
La$_{1-x}$Sr$_x$MnO$_3$ $\sigma_0^{inc}\sim
0.4\cdot 10^3\;(0.3\cdot 10^3)\; (\Omega$cm)$^{-1}$ 
for the doping concentrations $x=0.175\;(0.3)$, respectively, at $T=10$K.

Hence we conclude that besides the energy scale
also the absolute value of the incoherent part of the experimental
$\sigma(\omega)$ spectrum is consistent with the orbital model.
We stress that the value (order of magnitude) of $\sigma_0^{inc}$ is essentially
fixed by the conductivity sum rule and the scale $\omega_{1/2}$, as long as
the conductivity is predominantly incoherent.
The insets in Fig.\ref{fig1} and \ref{fig2} show $\sigma_0(\omega)$, Eq.(\ref{eq:sigma0}), 
with the Drude absorption 
at low frequency included, where we used an ad hoc chosen parameter
$\Gamma=0.1 t$ to broaden the $\delta$-function. 
This value corresponds to the experimental width $\Gamma\sim 0.02$ eV
taken from the (small) Drude peak observed by Okimoto {\it et al.}\cite{Okimoto97}.
While our model yields the weight of the Drude peak $D_c$, it does not give
$\Gamma$, which is due to extrinsic processes (e.g. scattering from impurities and grain
boundaries).
The charge stiffness $D_c$ together with $\Gamma$  determines
the DC-conductivity. 

Adopting the experimental estimate for $\Gamma$ we find for the example
in Fig.\ref{fig1} (inset)  for the low temperature DC-conductivity
$\sigma_{DC}\sim 6.5\cdot 10^3\;(\Omega$cm)$^{-1}$ and for the resistivity
$\rho\sim 0.15\cdot 10^{-3}\;\Omega$cm, respectively. 
For comparison, the experimental range of 
resistivities is e.g.  $0.1-1.0\cdot 10^{-3}\;\Omega$cm in the ferromagnetic
metallic phase of  La$_{1-x}$Sr$_x$MnO$_3$\cite{Urushibara95} . 
As we shall see below, the Drude weight and therefore the DC-conductivity 
depend considerably on the value of the exchange coupling $J$, and whether the
3-site hopping processes in the model are taken into account or not.
These terms have a strong effect on the coherent motion of charge
carriers. After this discussion it should be clear, however, that the 
DC-conductivity at low temperatures is largely determined by extrinsic effects.
\begin{figure}
\epsfxsize=6.6cm
\centerline{\epsffile[60 20 530 680]{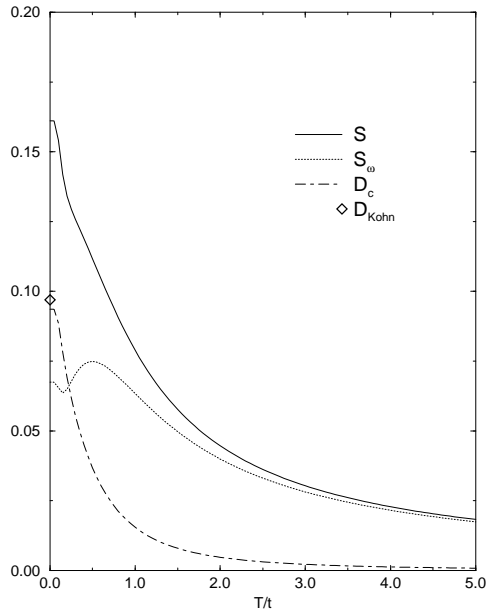} }
\noindent
\caption{ \label{fig3}
Temperature dependence of the kinetic energy S (solid line), 
the incoherent spectral weight $S_{\omega}$ (dotted) and the
Drude weights $D_{c} $ (dash-dotted line) and $D_{Kohn} $ (diamond)
for a N=10 site plane with $ N_e $=8  electrons and J=0.25t (U=16t)
in units of $t/\rho_0 e^2$. }
\end{figure}

Figure \ref{fig3} shows the temperature variation of the sum rule (kinetic energy),
the finite frequency absorption $S_{\omega}$ and the Drude weight $D_c$ 
for the model including the 3-site terms.
Whereas at high temperature the sum rule is essentially exhausted by the 
finite frequency absorption $\sigma(\omega)$, we find at low temperatures a 
significant increase of the Drude weight. In this particular case $D_c$
contributes about 40$\%$ to the sum rule at $T/t=0.3$.

The conductivity data shown in Figs. \ref{fig1} and \ref{fig2} 
for $T/t>0.5$ is characteristic for a 2D orbital 
liquid state, which is stabilized by thermal fluctuations.
The significant increase of coherency in Fig. \ref{fig3} below $T/t=0.3$ is due to
the onset of $x^2$-$y^2$ orbital order in the planar model (see discussion below).
This ordering is characteristic for the 2D
version of the orbital $t$-$J$ model and does probably not occur in 
cubic systems for which an orbital liquid ground state was proposed\cite{Ishihara97b}.
Therefore we consider the $T/t=0.3$ data for $D_c$ to be more appropriate
for a comparison with the low-temperature 3D data than the $T=0$ values,
which are enhanced due to orbital order.

We have also calculated $D_c$ for $T=0$ using Kohn's relation
\cite{Kohn64}
\begin{equation}
D_{Kohn}=\frac{1}{2N}\frac{\partial^2 E_0(\Phi)}{\partial\Phi^2}
\biggr|_{\Phi=\Phi_0},
\label{D_Kohn}
\end{equation}
which yields a consistent value (Fig.\ref{fig3}). 
Here $\Phi=eaA_{\alpha}/\hbar$ is the Peierls 
phase induced by the applied vector potential with component $A_{\alpha}$.
For the evaluation of Eq.\ref{D_Kohn} we followed Ref.\cite{Stephan92} and first searched
for the minimum of the ground state energy $E_0$ with respect to the
vector potential\cite{BC}, and then calculated the curvature. 
Since this procedure is quite cumbersome, we have determined most of our
$D_c$ data via the sum rule. 
The resulting sets of data show systematic trends as function of $J$ and
doping.

From these results we expect in analogy with the $t$-$J$ model that the
single-particle electron Green's function is characterized by a pronounced
quasiparticle peak to explain such a large fraction of coherent transport.

From Fig.\ref{fig3} we also see that the weight under $\sigma(\omega)$, i.e. $S_{\omega}$,
does not further increase for temperatures below $T=0.5 t$, although the
temperature variation in Fig.\ref{fig1} seems to suggest a significant further increase
towards lower temperatures. Conductivity data for $T<0.5 t$ is not shown
in Figs. \ref{fig1} and \ref{fig2} because it shows pronounced discrete level structure, 
and probably requires larger clusters for a careful study.
Nevertheless it appears that  $\sigma(\omega)$ develops a pseudogap in the orbital
ordered phase\cite{2D}.
Integrated quantities on the other hand are much less influenced by such
effects.\newline 
\newline\noindent
{\bf The Role of 3-Site Terms}\newline

\begin{figure}
\epsfxsize=6.6cm
\centerline{\epsffile[60 20 530 680]{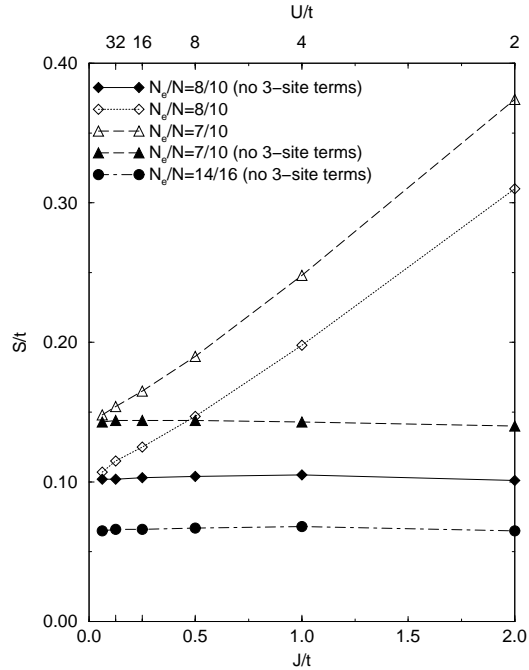} }
\noindent
\caption{ \label{fig4}
Kinetic energy S versus J for a N=10 site 2D cluster with $ N_e $=7 and 8 electrons,
with (full) and without (open symbols) the 3-site terms ($T/t =0.3$), respectively.
The dash-dotted line (circles) represents a 16 site
cluster with  $ N_e $=14 and without 3-site terms (From Horsch, Jakli\v c and 
Mack\cite{Horsch98}).
}
\end{figure}

\begin{figure}
\epsfxsize=6.6cm
\centerline{\epsffile[60 20 530 680]{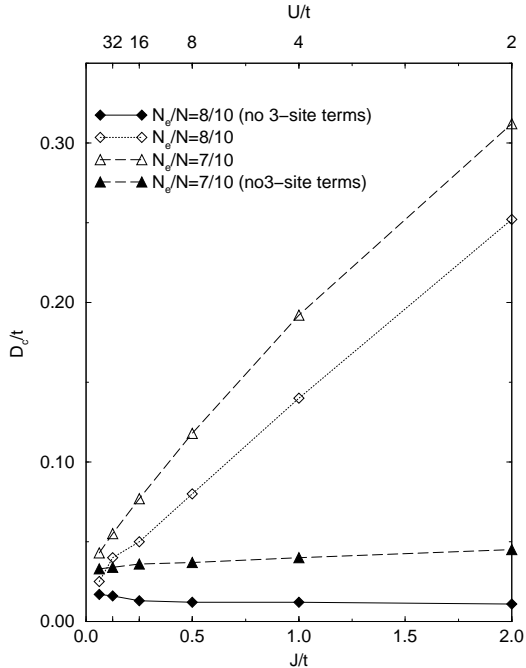} }
\noindent
\caption{ \label{fig5}
Drude weight $D_{c}$ versus $J$ calculated via the sum-rule (\ref{eq:Dc}) for a 
N=10 site planar cluster with $ N_e $=7 and 8 $e_g$ electrons,
with and without the 3-site terms, respectively. 
}
\end{figure}

In the following we wish to shed more light on the role of the 3-site
processes in the orbital Hamiltonian and in the current operator. 
The importance of the 3-site term becomes particularly clear from the $J$-dependence
of the sum rule $S$ (Fig.\ref{fig4}) and the charge stiffness $D_c$ (Fig.\ref{fig5})
taken at temperature $T/t=0.3$.
These low temperature values of $S$ and $D_c$ are approximatively independent of
the exchange parameter $J$ in the absence of 3-site terms. 
A similar observation was made  for the $t$-$J$ model
\cite{Poilblanc91,Stephan92}. Moreover one can see that
the sum rule $S$ is proportional to the doping concentration $x$
for the 3 cases shown  ($x$=0.125, 0.2, and 0.3).
When 3-site terms
are taken into account both $S$ and $D_c$ acquire a component which increases
linearly with $J$. These general features are fully consistent with results 
for the $t$-$J$ model\cite{Stephan92}.

\vspace{4mm}
%
%

The change due to the 3-site terms is particularly large for the charge 
stiffness, which defines the weight of the low frequency Drude peak.
The relative weight in the Drude peak $D_c/S$ is shown in Fig. \ref{fig5} as function
of $J=4 t^2/U$. Small changes in $J$ lead to considerable changes in the Drude
weight and the coherent motion of carriers. 
For $J=0.25\;(0.5)$ the 3-site terms lead to an increase of $D_c$ by a factor
of 3 (5), respectively, for the 8 electron case.

When we take the value $t\sim 0.2$ eV, which we determined from the comparison of 
$\sigma(\omega)$ with the experimental data, and $U\sim 3$ eV\cite{Ishihara97a}, 
we obtain $U/t\sim 15$
and $J/t=4 t/U \sim 0.25$ ($J=0.05$ eV). For such a value for $J$ we expect a relative
Drude weight $D_c/S \sim 0.4$, which is larger than the experimental
ratio $D_c/N_{eff}\sim 0.2$ found by Okimoto {\it et al.} for $x=0.175$, identifying
here the effective number of carriers $N_{eff}$\cite{Okimoto97} with $S$. 
The small value for  $D_c/N_{eff}$ found in the experiments suggests that $J$
is not larger than the value estimated, otherwise we would expect a 
too large relative Drude weight. A larger value for $U$ would also have 
the effect to reduce $D_c$.
However, we have to stress here that certainly calculations on larger clusters 
must be performed, before possible finite size effects in $D_c$ can be quantified.
\newline
\newline\noindent
{\bf Comparison with the standard $t$-$J$ Model}\newline

\begin{figure}
\epsfxsize=6.6cm
\centerline{\epsffile[60 20 530 680]{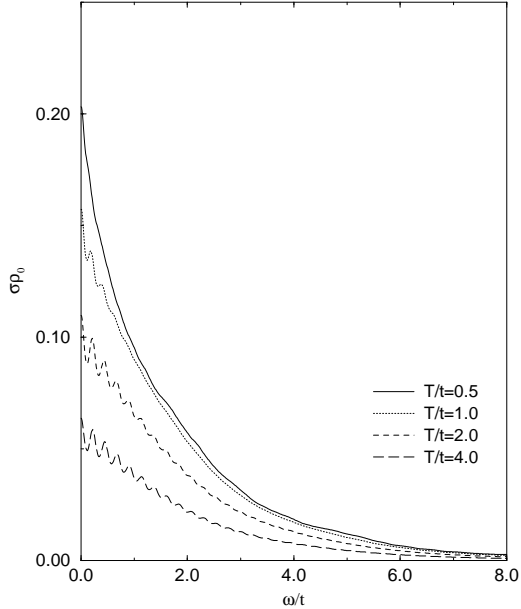}}
\noindent
\caption{ \label{tj1}
Optical conductivity for the {\em standard $t$-$J$ model}
($t_{aa}=t_{bb}=t$ and $t_{ab}=0$) at temperatures
$T=0.5, 1.0, 2.0 $ and $4.0$ which should be compared with the corresponding 
data for the orbital model. Here $N_{h}/N=3/16$ and $J/t=0.4$.
}
\end{figure}

\begin{figure}
\epsfxsize=6.6cm
\centerline{\epsffile[60 20 530 680]{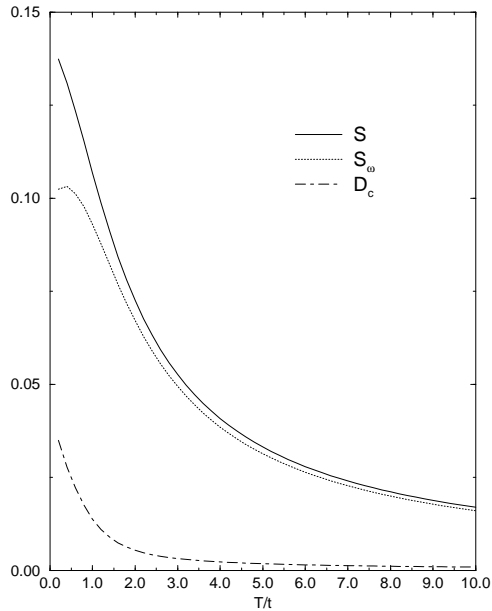}}
\noindent
\caption{ \label{tj2}
Temperature dependence of the kinetic energy S (solid line), 
the incoherent spectral weight $S_{\omega}$ (dotted) and the
Drude weight $D_{c} $ (dash-dotted line) for the {\em $t$-$J$ model}.
The data was calculated for a 2D system with 16 sites and 3 holes
for $J/t$=0.4.
}
\end{figure}

It is interesting to compare these results for the orbital $t$-$J$ model
with those obtained for the usual $t$-$J$ model describing the carrier motion
in the copper-oxygen planes in high-T$_c$ superconductors. 
We note that 
the $t$-$J$ model is a special case of the orbital model with $t_{aa}=t_{bb}$
and $t_{ab}=0$.
Figures \ref{tj1} and \ref{tj2} show results for a $4\times 4$  cluster with
$N_h=3$ holes. 
The results for $\sigma(\omega)$ show for both models a
large incoherent absorption, which is bell-shaped in the case of the orbital 
model, while in the $t$-$J$ case there is a continuous increase of 
$\sigma(\omega)$ as $\omega$ approaches zero.
The temperature dependence of the sum rule, the incoherent and the Drude
weight shown in Fig. \ref{tj2} are quite similar for both models, with
the important exception that the orbital model goes into the ordered phase
at low temperatures ($T\leq 0.3$) which is accompanied by a marked increase of
$S$ and $D_c$ (Fig.\ref{fig3}). A more detailed study of the $T$-dependence
may be found in the review by Jakli\v c and Prelov\v sek\cite{Jaklic98a}.
\newline

\vspace*{-2cm}
%


\vspace*{2cm}\noindent
{\bf Comparison with other Work}\newline

The optical conductivity for the orbital non-degenerate
Kondo lattice model was studied by Furukawa
using the dynamical mean-field approximation
\cite{Furukawa95}. An
important result of this calculation is, that the weight of intraband 
excitations within the lower exchange-split band should be proportional 
to the normalized ferromagnetic magnetization $M/M_{sat}$. Experimentally, however,
$N_{eff} (\sim S)$ is still increasing when temperature is lowered even
though $M$ is already saturated. This contradiction to the prediction
of the simple double-exchange model implies that some other large-energy-scale
scattering mechanism survives at low temperature, where the spins are frozen
\cite{Okimoto97}. The dynamical mean field theory yields one low energy scale,
and not two, i.e. there is no Drude peak plus incoherent structure.

Shiba and coworkers {\it et al.}\cite{Shiba97} approached the problem from the
band picture.
They analysed the noninteracting two-band
model (2) (i.e. without constraint) and argued that the interband transitions 
within the $e_g$ orbitals may explain the anomalous absorption in the ferromagnetic 
phase of La$_{1-x}$Sr$_x$MnO$_3$.
While the frequency range of the interband transition is found consistent 
with the anomalous absorption, the structure of $\sigma(\omega)$ differs.
In particular the noninteracting model leads to $\sigma(\omega)\sim \omega$
for small $\omega$.
Moreover they calculated the ratio
$S_{\omega}/D_c \sim 0.85$, i.e. there is more weight in the Drude peak than in the 
regular part of the conductivity $\sigma(\omega)$. This ratio is found to be
rather insensitive to doping in the range $0.175<x<0.3$.
This result differs considerably from the experimental data of Okimoto {\it et al.}
where this ratio is about 4 for $x=0.175$.

Orbital fluctuations in the ferromagnetic state and their effect 
on the optical conductivity were studied by Ishihara {\it et al.}
\cite{Ishihara97b} using slave fermions coupled to bosonic orbital
excitations.  Ishihara {\it et al.} deal with the concentration regime $xt \gg J$, where the 
kinetic energy is expected to dominate the orbital exchange energy,
and arrive at the conclusion that the quasi two-dimensional 
nature of the orbital fluctuations leads to an {\em orbital liquid} in (3D) cubic systems.
The orbital disorder is treated in a static approximation in their work.
The $\sigma(\omega)$ spectrum obtained by this approach for $T=0.1 t$
has a quite similar shape as the experimental spectrum at low temperature
(with $\omega_{1/2}\sim 3 t$), however, because of the assumed static disorder,
there is no Drude component.
A further problem with the orbital liquid, as noted by the authors, is the
large entropy expected for the orbital disordered state which seems to be 
in conflict with specific heat measurements\cite{Woodfield97}.

More recently a microscopic theory of the optical conductivity based on a slave boson
parametrization which combines strong correlations and orbital degeneracy
was developed by Kilian and Khaliullin\cite{Kilian98}
for the ferromagnetic state at zero temperature. Their
approach yields a highly incoherent spectrum due to the scattering
of charge carriers from dynamical orbital fluctuations, and a Drude 
peak with strongly reduced weight. 
However $\sigma(\omega)$ is depressed at low frequency
in this calculation for the
orbital model and an additional electron-phonon mechanism is invoked
to obtain results similar to the experimental spectral distribution.
This theory further accounts for the small
values of the specific heat and therefore supports the orbital liquid
scenario for the ferromagnetic cubic systems\cite{Ishihara97b}. 

We would like to stress that the proposed orbital liquid state\cite{Ishihara97b} 
is a property of cubic
systems, and not a property of the 2D or quasi-2D versions of the
orbital model. In the planar model the cubic symmetry is broken from
the outset and holes presumably cannot restore it.
This is different from the physics of the $t$-$J$ model, where holes 
restore the spin rotational symmetry (which is spontaneously broken 
in the antiferromagnetic ordered phase), because this symmetry is 
respected by the $t$-$J$ Hamiltonian in any dimension.

For a deeper understanding of the results for the optical conductivity,
we shall analyse below the structure of the orbital
correlations in the doped and undoped two-dimensional orbital model
in detail.

\vspace{1.1cm}\noindent
{\bf DOPING DEPENDENCE OF ORBITAL CORRELATIONS}\newline

In the absence of holes the  interaction  (\ref{eq:H_orb})
between orbital pseudo-spins
will lead to an orbital order below a certain temperature $\sim J$
due to the anisotropy of the Hamiltonian.
Doping will destroy this order and lead either to a disordered
orbital liquid, 
or may generate a new kind
of ordered state which optimizes the kinetic energy of the holes.
To investigate this question for an anisotropic model 
it is in general not sufficient to calculate simply
e.g. the correlation function $<T^z_{\bf i}T^z_{\bf j}>$ defined with respect
to the original orbital basis,
because the relevant occupied and unoccupied orbitals may be
different from the original orbital basis chosen.
In the following we therefore introduce a local orthogonal transformation of the orbitals 
on different sublattices by angles $\phi$ and $\psi$, respectively. 
On the A-sublattice:
\begin{eqnarray}
|{\tilde z}\rangle &=& cos(\phi)|z\rangle + sin(\phi)|x\rangle \nonumber\\
|{\tilde x}\rangle &=& -sin(\phi)|z\rangle + cos(\phi)|x\rangle , 
\end{eqnarray}
and a similar rotation with angle  $\psi$ on the B-sublattice.
This amounts to new operators, e.g.
\begin{eqnarray}
{\tilde T}^z_i&=&cos(2\phi) T^z_i+sin(2\phi)T^x_i \\
{\tilde T}^z_{i+R}&=&cos(2\psi) T^z_{i+R}+sin(2\psi)T^x_{i+R}
\end{eqnarray} 
and new correlation functions, e.g.  
$\langle{\tilde T}^z_{\bf i}{\tilde T}^z_{\bf j}\rangle$ is defined as
\begin{eqnarray}
\langle {\tilde T}^z_i{\tilde T}^z_{i+R}\rangle=
cos(2\phi)cos(2\psi)\langle   T^z_i T^z_{i+R} \rangle 
+sin(2\phi)sin(2\psi)\langle  T^x_i T^x_{i+R}\rangle \nonumber \\
+cos(2\phi)sin(2\psi)\langle  T^z_i T^x_{i+R}\rangle 
+sin(2\phi)cos(2\psi)\langle  T^x_i T^z_{i+R} \rangle
\end{eqnarray} 
The angles $\phi$ and $\psi$ are now chosen such that the 
nearest neighbor correlation function
$\langle{\tilde T}^z_{\bf i}{\tilde T}^z_{\bf j}\rangle$
takes a maximal (ferromagnetic) value. With this convention
the orbital order is specified by the corresponding local 
quantization axis, that is by the  values of 
the rotation angles $\phi$ and $\psi$. 
We call this the {\em optimized orbital basis} (OOB)\cite{Horsch98}.

\begin{figure}
\epsfxsize=6.6cm
\centerline{\epsffile[60 20 500 700]{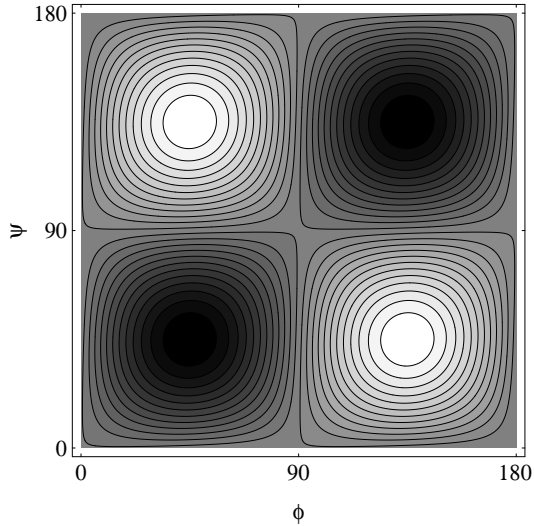} }
\noindent
{\vspace*{-2.3cm}\noindent
\caption{ \label{fig7}
Contourplot of the {\em  rotated} nearest neighbor orbital correlation function
$ \langle  {\tilde T}^z_i {\tilde T}^z_{i+R} \rangle$,  $R$=(1,0), as function of the 
angles $ \phi $ and $ \psi $ for a N=10 site planar cluster with $ N_e
$=10 electrons, J=0.25t (U=16t) and T=0.01t. White regions correspond to positive
(ferromagnetic) and black areas to negative (antiferromagnetic) correlation functions,
i.e. $ \langle  {\tilde T}^z_i {\tilde T}^z_{i+R} \rangle$$>$0.23 ($<$-0.23), 
respectively.
The 25 contour lines are chosen equidistant in the intervall [-1/4,+1/4].
(From Horsch, Jakli\v c and Mack\cite{Horsch98}) }}
\end{figure}
\begin{figure}
\epsfxsize=7.0cm
\centerline{\epsffile[60 0 530 680]{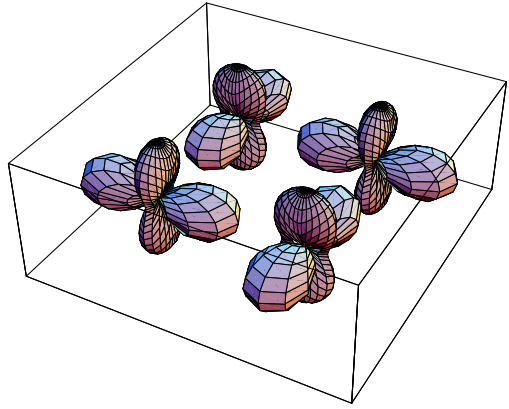} }
{\vspace*{-2.0cm}\noindent
\caption{ \label{fig13}
Alternating  $\frac{1}{\sqrt{2}}(|x\rangle + |z\rangle )$ and
$\frac{1}{\sqrt{2}}(|x\rangle - |z\rangle )$
orbital order in the undoped planar system.
}}
\end{figure}
\begin{figure}
\epsfxsize=6.6cm
\centerline{\epsffile[60 20 530 680]{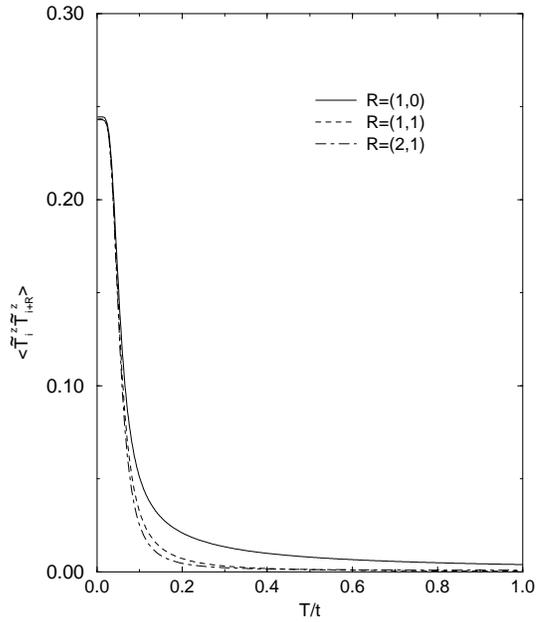} }
\noindent
\caption{ \label{fig8}
Temperature dependence of  $ \langle  {\tilde T}^z_i {\tilde T}^z_{i+R} \rangle
$ with $ \phi =45^0$ and $ \psi=135^0$ for nearest-neighbors and 
$ \psi=45^0$ for next-nearest neighbors, respectively. Results are shown for
different distances $R$ for an undoped $N=10$ site planar cluster.
Parameters as in Fig.\ref{fig7}. The strong increase for $T/t\leq 0.1$ is due to the onset
of alternating orbital order.  (From Horsch, Jaklic and Mack\cite{Horsch98})
}
\end{figure}

In the undoped 2D case (x-y plane) these angles are $\phi =45^o$ and
$\psi =135^o$ corresponding to an $\frac{1}{\sqrt{2}}(|x\rangle + |z\rangle )
$ and  $\frac{1}{\sqrt{2}}(|x\rangle - |z\rangle) $ order at low temperatures
(see Fig. \ref{fig7} and  \ref{fig13}). 
$ \langle {\tilde T}^z_i {\tilde T}^z_{i+R} \rangle $ takes a
value only slightly smaller than 0.25 at low temperature (Fig. \ref{fig8}),
which implies that quantum fluctuations are small in this state. 
Hence the orbital correlation function  
$ \langle  {\tilde T}^z_i{\tilde T}^z_{i+R} \rangle$  of the undoped orbital
model is {\em alternating} (i.e. antiferromagnetic in the pseudo-spin language
for the orbital degrees of freedom). 
That the AF-order seems to be established at a finite
temperature for the small clusters means that the correlation length
$\xi$ gets larger than the system size L.
At small temperatures the correlation functions have about the same value 
independent of $R$, i.e. the data shows no spatial decay.
Above $T/t=0.1$ the orbital correlations are small and show a pronounced 
spatial decay, which can be considered as a signature of a thermally disordered 
orbital liquid state.

Experimentally the coexistence of
antiferromagnetic (or staggered) orbital order 
and ferromagnetic spin order has been
established in the low-doping regime of LaMnO$_3$\cite{Goodenough55,Goodenough71}.
It should be recalled that  LaMnO$_3$ has an A-type antiferromagnetic spin structure
\cite{Wollan55}, where ferromagnetic layers are coupled antiferromagnetically
along the $c$-axis.
In the orbital ordered case the MnO$_6$ octahedra are deformed, and Mn $3d_{3x^2-r^2}$
and $3d_{3y^2-r^2}$ orbitals with orientation along the $x$- and $y$-axis,
respectively, have been considered as relevant occupied orbitals
\cite{Matsumoto70,Elemans71}.
For the planar model we find here the alternate occupation of 
$\frac{1}{\sqrt{2}}(|x\rangle - |z\rangle )$ and 
$\frac{1}{\sqrt{2}}(|x\rangle + |z\rangle )$ orbitals (Fig. \ref{fig13}) 
which are also oriented along
the $x$- and $y$-axis, however they differ with respect to their shape
(and have pronounced lobes along $z$).\\

\vspace*{-1.0cm}.

\begin{figure}
\epsfxsize=6.6cm
\centerline{\epsffile[60 20 500 700]{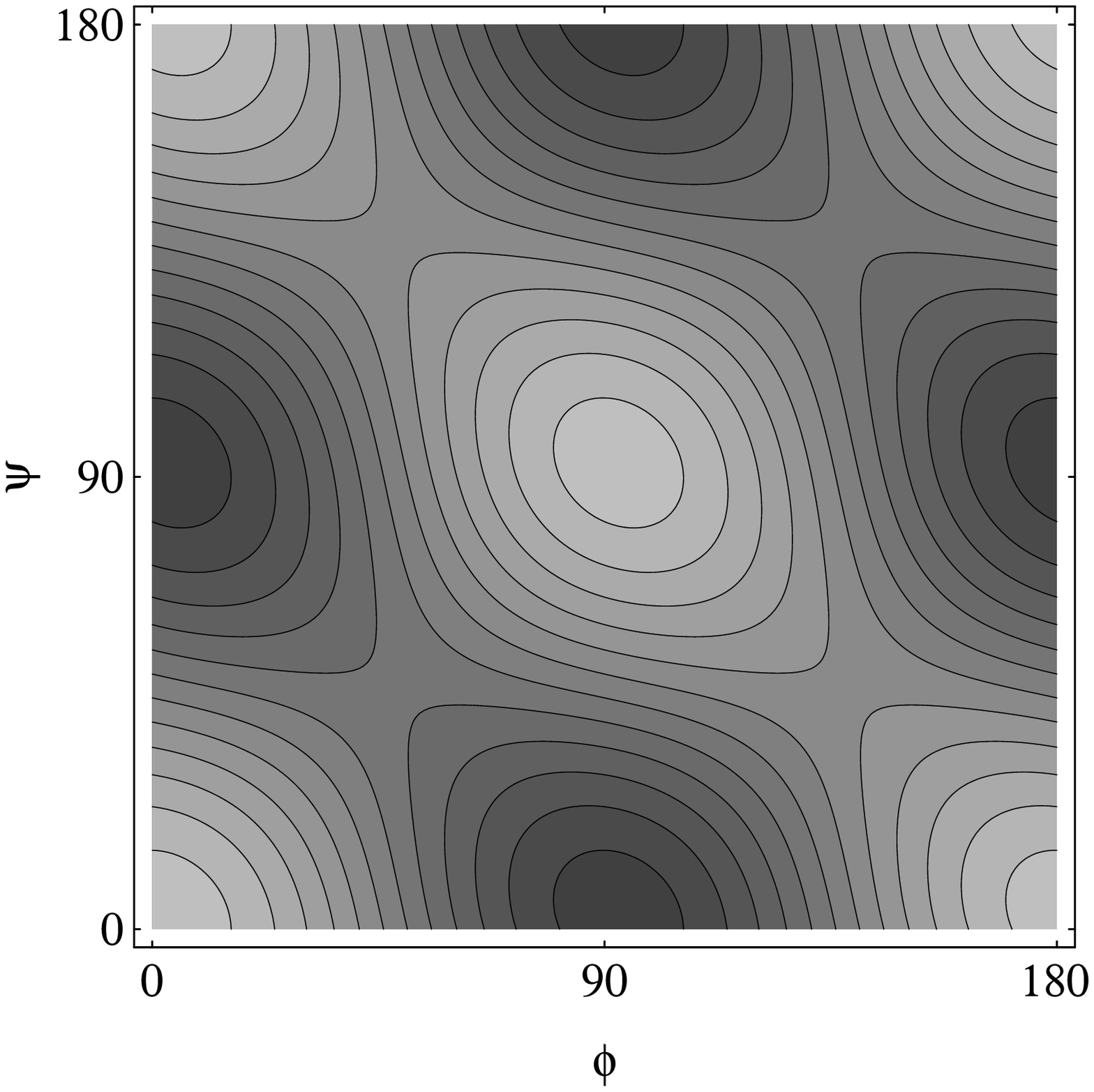} }
\noindent
\caption{ \label{fig9}
Contourplot of $ \langle {\tilde T}^z_i {\tilde T}^z_{i+R} \rangle$ for $R=(1,0)$
as function of angles $ \phi $ and $ \psi $ 
for a N=10 site cluster with two holes. Otherwise same parameters as for Fig.\ref{fig7}.
}
\end{figure}

\begin{figure}
\epsfxsize=7.0cm
\centerline{\epsffile[60 20 530 680]{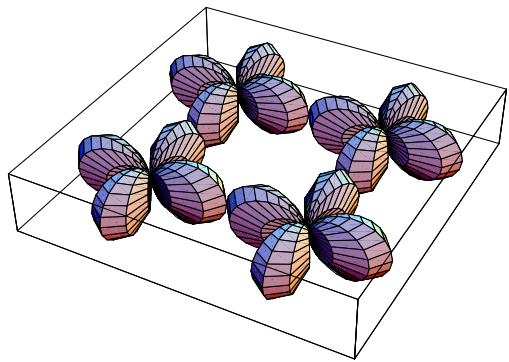} }
{\vspace*{-2.0cm}\noindent
\caption{\label{fig14}
$x^2$-$y^2$ orbital order in the doped phase.
}}
\end{figure}

\begin{figure}
\epsfxsize=6.6cm
\centerline{\epsffile[60 20 530 680]{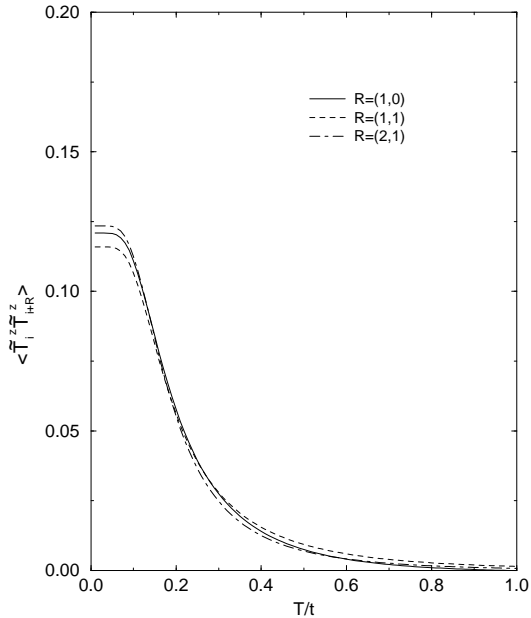} }
\noindent
\caption{ \label{fig10}
Temperature dependence of $ \langle {\tilde T}^z_i {\tilde T}^z_{i+R} \rangle$ 
with $ \phi =0^o $ and $ \psi=0^o $ for a 10-site cluster with two $e_g$ holes.
Parameters as in Fig.\ref{fig7}.
The strong increase below $T/t=0.3$ signals the onset of $x^2$-$y^2$ orbital
order in the planar model.
}
\end{figure}

Doping of two holes in a ten site cluster is 
sufficient to remove the alternating orbital order
and establishes a ferromagnetic orbital order ($\phi = \psi =0$) 
as can be seen from Figs. \ref{fig9} and \ref{fig10}. 
This corresponds to preferential occupation of orbitals with symmetry
$x^2-y^2$ as shown in Fig.\ref{fig14}.
The pseudospin alignment can be considered as a kind of 
{\em double exchange mechanism in the orbital sector}.
An interesting feature in Fig.\ref{fig10} for the doped case is the fact that
the correlations do not show significant spatial decay even at higher 
temperatures where the correlations are small.
Moreover the correlations in the doped case appear to be more robust
against thermal fluctuations than in the undoped case.
The characteristic temperature (determined from the half-width in Figs.\ref{fig8} 
and \ref{fig10})
is $T^*\sim 0.2 t$ for the 2-hole case while for the undoped system $T^*\sim 0.05 t$.
This trend is consistent with the fact that in the doped case
order is induced via the kinetic energy and the corresponding scale $t$
is larger than $J=0.25 t$. Although a more detailed analysis would be 
necessary to account properly for the doping dependence.

It has been shown recently by Ishihara {\it et al.}\cite{Ishihara98c} that in
the layered manganite compounds hydrostatic pressure leads to a stabilization of the 
$x^2$-$y^2$ orbitals as well.

\begin{figure}
\epsfxsize=7.6cm
\centerline{\epsffile[60 20 530 680]{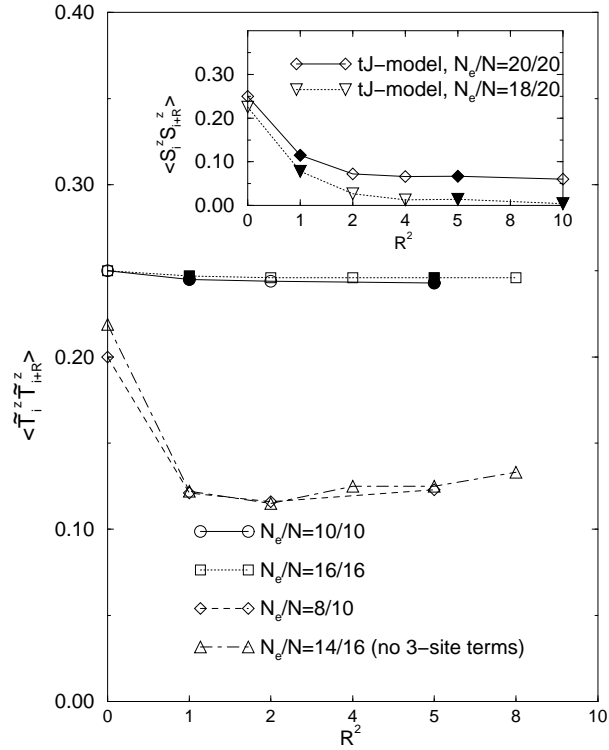} }
\noindent
\caption{ \label{fig11}
Orbital correlation functions for $T=0$
as function of distance (squared)
for a 10-site and a $4\times 4$ cluster with zero and two
holes ($J/t=0.25$). 
In the undoped case the correlation function is antiferromagnetic,
whereas in the doped case orbital correlations are ferromagnetic. 
Antiferromagnetic orbital order is indicated by full (negative) and 
open (positive) symbols, respectively. 
Inset: The spin correlations $\langle S^z_i S^z_{i+R}\rangle$
for a 20-site 2D Heisenberg model and
the $t$-$J$ model with two holes are shown for comparison 
($t$-$J$ data for $J=0.4$ taken from Horsch and Stephan\cite{Horsch91}). 
The two-hole case shows a rapid decay of correlations
characteristic for an antiferromagnetic spin liquid.
}
\end{figure}

A comparison of the doping dependence of spatial correlations 
with the $t$-$J$ model is given in Fig.\ref{fig11}.
In the undoped case the orbital model shows almost classical (orbital)
N\'eel order, whereas in the $t$-$J$ model correlations are strongly reduced
by quantum fluctuations. The $T=0$ correlation function of the $t$-$J$
model shows long-range order (LRO) consistent with a strongly reduced
sublattice magnetization $m_z\sim 0.3$ (in the thermodynamic limit).

In the orbital model the effect of the two holes ($x=0.125$ and $0.2$) is 
strong enough to induce
ferromagnetic (orbital) correlations, i.e. with a prefered occupation
of $x^2$-$y^2$ orbitals.
In the $t$-$J$ model instead spin-correlations decay rapidly for two holes on
a 20-site cluster ($x=0.1$), 
which is consistent with the notion of an AF spin-liquid state 
(see inset Fig.\ref{fig11}).

The preference of ferromagnetic orbital order in the doped case is due to the
fact that $t^{\alpha \beta}$ depends on the orbital orientation, 
and $t^{aa}\gg t^{bb}$.
One would expect that for sufficiently strong orbital exchange 
interaction $J=4 t^2/U$ one reaches a point where antiferromagnetic orbital
interactions and ferromagnetic correlations due to the kinetic energy are 
in balance. This quantum critical point between FM and AF orbital order turns 
out to be at quite large $J$ values. For the two-hole case 
($x=0.2$) we have found that this crossover happens for quite large orbital 
interaction $J\sim 2$, i.e. at a value about an order of magnitude larger 
than typical $J$ values in manganite systems.

Before closing this section, we shall explain the different orbital occupancy
in the doped and undoped case in more physical terms.
The overlap of the two sets of orthogonal $e_g$ orbitals 
on neighboring sites changes with the angles $\phi$ and $\psi$.
(1) In the undoped state the direct hopping between the (predominantely) occupied
orbitals  $\frac{1}{\sqrt{2}}(|x\rangle + |z\rangle )$
and  $\frac{1}{\sqrt{2}}(|x\rangle - |z\rangle) $  on neighboring sites
is small (Fig.\ref{fig13}). The hopping between these occupied
orbitals is blocked anyhow because of Pauli's principle.
However $t^{ab}$ between an occupied orbital on one site and an unoccupied
orbital on a neighbor site is extremal. This leads to a maximal antiferromagnetic
orbital exchange interaction, and thereby to a lowering of the energy.
(2) In the doped case instead, the $x^2$-$y^2$ orbital occupancy is prefered on all
sites (Fig.\ref{fig14}), which implies a large hopping amplitude for holes in the 
partially occupied $x^2$-$y^2$
band, whereas the off-diagonal hopping matrix element $t^{ab}$ and
the orbital interactions are reduced for this choice of occupied orbitals.
Hence the main energy gain in this case is due to the  
correlated motion of holes in the $x^2$-$y^2$-band.
Nevertheless the off-diagonal hopping plays an important role in the 
$x^2$-$y^2$ ordered state, and leads to the strong incoherent features 
characteristic for the conductivity $\sigma(\omega)$.

Finally we note that
recent experiments by Akimoto {\it et al.}\cite{Akimoto98}
proved the existence of an A-type antiferromagnetic metallic ground
state in a wide doping concentration regime of the 3D system 
(La$_{1-z}$Nd$_z$)$_{1-x}$Sr$_x$MnO$_3$ and in particular for
La$_{1-x}$Sr$_x$Mn O$_3$ at high doping $x=0.52-0.58$.
The latter compound was investigated by Okimoto {\it et al.} for smaller
doping $x$ where the ground state is the uniform ferromagnetic state.
The A-type antiferromagnet consists of ferromagnetic layers which
are coupled antiferromagnetically. 
Akimoto {\it et al.} propose that $x^2-y^2$ orbitals should be occupied in 
this state and form a highly anisotropic 2D band. 
This seems consistent with our finding of  $x^2-y^2$ orbital order 
in the ferromagnetic planar system where the cubic symmetry is
explicitely broken.
An experimental study of the optical properties in this concentration
regime at low temperature would be particularly interesting.

%
%
%

\vspace{1.1cm}\noindent
{\bf SUMMARY AND CONCLUSIONS}\newline

Our work was motivated by the experimental study of the frequency dependent
conductivity of La$_{1-x}$Sr$_x$MnO$_3$    by Tokura's group\cite{Okimoto95}.
In particular these experiments show in the saturated low-$T$ ferromagnetic
phase a broad incoherent spectrum and at small frequency in addition a narrow
Drude peak, which contains only about 20$\%$ of the total spectral weight.
This experiment by itself shows that {\em the ferromagnetic metallic state
(although fully spin polarized) is highly anomalous }. 
These features cannot be explained by the standard double exchange model,
but require an additional mechanism.
The most natural mechanism, one can think of, is the twofold degeneracy
of the $e_g$ orbitals.

We have studied the optical conductivity, the charge stiffness and the optical
sum rule for an effective Hamiltonian that contains only the $e_g$-orbital
degrees of freedom. This model follows from the more general Kondo lattice
model, when the spins are ferromagnetically aligned.
The `orbital $t$-$J$ model' is expected to describe the low-temperature physics
of manganese oxides in the ferromagnetic saturated phase.

Our results for $\sigma(\omega)$ 
in the 2D orbital disordered regime
show that these characteristic features 
follow from the orbital model.
We believe that the optical conductivity in the 2D orbital liquid regime ($T>T^*$)
can be compared with the conductivity in the 3D-orbital liquid state. 
In particular there is (1) a broad continuum which decreases towards 
high frequency with a half-width $\omega_{1/2} \sim 2.5$ and $3.5 t$ 
in two and three dimensions, respectively.
(2) The absorption increases with decreasing temperature and (3) the
absolute value of the incoherent part of $\sigma(\omega)$ at low
frequency and temperature $\sigma^{inc}_0 \sim 0.4 \cdot 10^3 (\Omega$cm)$^{-1}$
has the correct order of magnitude as in Tokura's experiments.
(4) Despite this strong incoherence there is a finite Drude peak of
about 20$\%$ of the total sum rule for small $J$.
(5) The coherent motion and the Drude peak become more pronounced for larger 
values of $J$ due to the 3-site hopping processes in the model.
For the value $J/t=0.25$ estimated here the relative Drude weight is 40 (20) $\%$
in the model with (without) 3-site hopping processes, respectively.
(6) The low temperature values for the DC-conductivity and the resistivity
can be estimated by assuming a width $\Gamma$ for the Drude peak (taken
from experiment).  The additional scattering processes contributing to
$\Gamma$ are due to impurity or grain boundary scattering, i.e. these 
are extrinsic scattering processes which are not contained in the
orbital model. The important information which follows from the model is 
the weight of the Drude peak, which together with $\Gamma$ determines
the DC-conductivity. 

The relatively small Drude peak in comparison with the incoherent part of 
$\sigma(\omega)$ clearly indicates that the carrier motion is essentially
incoherent. Nevertheless we expect a quasiparticle peak in the single particle
Green's function, yet with small spectral weight.
Our study of the frequency dependent conductivity shows that the (saturated)
ferromagnetic state in the manganites has unconventional transport properties
due to the scattering from orbital excitations in combination with the
exclusion of double occupancy.  
A detailed study of the orbital dynamics in the doped system is necessary
for a deeper understanding of these issues.

The orbital model resembles the $t$-$J$ model, which describes the correlated motion
of charge carriers in the cuprate superconductors. 
Yet an important difference is the cubic symmetry of the pseudospin representation
of the orbital degrees of freedom.
The hopping matrix elements depend on the orbital orientation, and are in
general different in the two diagonal hopping channels. Moreover there is
also an off-diagonal hopping matrix element, which does not exist at all
in the $t$-$J$ model.
At low temperatures ($T<T^*\sim 0.2 t$)
doping induces (ferromagnetic) $x^2$-$y^2$ {\em orbital order} for realistic values for the
exchange interaction $J$ in the 2D orbital model.
This is in striking contrast to $t$-$J$ physics in cuprates, 
where the system changes upon doping
from antiferromagnetic long-range order to an antiferromagnetic spin liquid.
In both models coherent motion coexists with strong incoherent features. 

Our calculations have shown that the orbital mechanism can explain the order of
magnitude of the conductivity at low temperature. 
The orbital degrees of freedom are certainly also important for a quantitative
calculation of the collossal magnetoresistance. This is obvious because in the full 
Kondo lattice model there is a close interplay between orbital  and
spin degrees of freedom\cite{Ishihara97a,Feiner98}.

We acknowledge helpful discussions with F. Assaad, J. van den Brink, L. Hedin,
G. Khaliullin, A. Muramatsu, A. M. Ole\'s and R. Zeyher.


\begin{thebibliography}{99}
\bibitem{BC}
The same twisted boundary conditions were used for the finite temperature studies
in this work. These boundary conditions guarantee a nondegenerate closed-shell
ground state.
\end{thebibliography}
\end{document}